# A novel super-elastic carbon nanofiber with cup-stacked carbon nanocones and a screw dislocation

Xu Han[1], Futian Xu[1], Shuyong Duan[1,*], Haifei Zhan[2,**], Yuantong Gu[2], and Guirong Liu [3,4]

[1]*State Key Laboratory of Reliability and Intelligence of Electrical Equipment, Hebei University of Technology, Tianjin, 300401 China*
[2]*School of Chemistry, Physics and Mechanical Engineering, Queensland University of Technology (QUT), Brisbane, QLD, 4001, Australia*
[3]*Department of Aerospace Engineering and Engineering Mechanics, University of Cincinnati, Ohio, USA*
[4]*Summer Part-Time Professor, School of Mechanical Engineering, Hebei University of Technology, Tianjin, 300401 China*

**Abstract.** Carbon nanofibers (NFs) have been envisioned with broad promising applications, such as nanoscale actuators and energy storage medium. This work reports for the first-time super-elastic tensile characteristics of NFs constructed from a screw dislocation of carbon nanocones (NF-S). The NF-S exhibits three distinct elastic deformation stages under tensile, including an initial homogeneous deformation, delamination, and further stretch of covalent bonds. The delamination process endows the NF-S extraordinary tensile deformation capability, which is not accessible from its counterpart with a normal cup-stacked geometry. The failure of NF-S is governed by the inner edges of the nanocone due to the strain concentration, leading to a common failure force for NF-S with varying geometrical parameters. Strikingly, the delamination process is dominated by the inner radius and the apex angle of the nanocone. For a fixed apex angle, the yielding strain increases remarkably when the inner radius increases, which can exceed 1000%. It is also found that the screw dislocation allows the nanocones flattening and sliding during compression. This study provides a comprehensive understanding on the mechanical properties of NFs as constructed from carbon nanocones, which opens new avenues for novel applications, such as nanoscale actuators.

---

*Corresponding author. Email: duanshuyong@hebut.edu.cn (Shuyong Duan)

**Corresponding author. Email: zhan.haifei@qut.edu.au (Haifei Zhan)

# 1. Introduction

Due to their intriguing mechanical, chemical and physical properties, carbon nanofibers (NFs) have shown appealing applications, such as artificial muscles [1, 2], intelligent textile and structural composites [3], sensors and actuators (e.g., tensile actuation or torsion actuation) [4-6], soft robotics [7, 8], flexible electronics [9], energy storage medium [10, 11], energy harvester [12]. Different types of NFs have been synthesized from either a chemical vapor deposition (CVD) method or spinning of carbon precursor, such as the NF with twisted-bundle of carbon nanotubes (CNTs), the NF with multi-walled CNT core, and the NF with stacking of graphene layers [13].

Particularly, for the stacked-up-type NFs, the graphene layers can be perpendicular, inclined or even coiled along the fiber axis [14-18]. One of the most frequently studied stacked-up structure is based on the carbon nanocone, which is first reported by Ge and Sattle [19]. It is theoretically and experimentally confirmed that the carbon nanocone can exhibit five apex angles, which are 19.2°, 38.9°, 60°, 83.6°, and 112.9°, respectively [20]. Further study reveals that the stability of the structure is sensitive to the cone apex angle, and an increase in the conical angle results in a moderate improvement in the structural stability [21]. Extensive studies have explored the applications of carbon nanocones, such as the nanoprobes for scanning probe microscopy [22], high resolution magnetic force imaging [23], and mass sensing [24], $H_2$ adsorption [25-27], Ne adsorption [28], electron field emitters [29], nanoindentation tips [30], and nanofillers for thermoplastic composites [31]. Recently, researchers find that the tapered surface of the carbon nanocone can move water droplet spontaneously at over 100 m/s [32], and its surface curvature gradient can be used to guide the directional motion of water droplet [33] and nanoparticles [34].

Besides the applications, extensive efforts have also been devoted to investigate the thermal conductivity of the single-walled (or mono-layer) carbon nanocones [35] and their mechanical properties [36-42]. The studied nanocones either have an open end or closed end, and the tensile [43] and compressive behaviors (especially the buckling and post-buckling behaviors) [44-47] are the most extensively studied deformation scenarios. Most of these studies rely on atomistic simulations [43, 46, 48], while the multi-scale quasi-continuum approach [45, 49] and continuum spring-mass model [39] have also been adopted. Different geometrical parameters have been assessed, including the apex angle, the length, the top radius and the bottom radius. Comparing with the mono-layer carbon nanocones, few studies have investigated the mechanical properties of the NFs with stacked carbon nanocones. A recent study show that the tensile properties of the cup-stacked NFs can be significantly changed through thermal treatments, which introduces surface bonds between adjacent layers [50].

Overall, previous studies have focused on the cup-stacked NFs with the graphene layers inclined to the fiber axis, and the NFs with a screw dislocation of carbon nanocones (NF-S) have been rarely discussed. Since the adjacent layers are adhered to each through the van der Waals (vdW) interactions, the NFs exhibit a small yielding strain around 7% [50]. Our recent works show that a helicoid nanostructure constructed from a screw dislocation of 2D nanomaterials possesses a super-elastic tensile characteristic [51, 52]. Therefore, it is of great interests to known how the NF-S with a screw dislocation would behave under external loading. To this end, here we carry out a series of in silico studies to explore the tensile characteristics of NF-S. It is found that the NF-S possesses a super-elastic tensile property with the yielding strain reaching up to 1727%, more than two orders higher than that the NF with an ordinary cup-stacked structure.

## 2. Method

The tensile properties of the cup-stacked carbon nanofiber were assessed through large-scale molecular dynamic (MD) simulations. As shown in **Figure 1**, the carbon nanofiber was constructed based on a screw dislocation of open-tip carbon nanocone (NF-S). The carbon nanocone was generated from a fan-shaped graphene nanosheet (Figure 1a) [13], which can be described by three geometric parameters, namely the inner radius ($r$), outer radius ($R$), and the sector angle ($\alpha$). Due to the ortho-hexagonal crystalline lattice, there are only five sector angles, i.e., 60°, 120°, 180°, 240°, and 300°. Based on the open-tip nanocone, the height of NF-S was determined by the layer number ($N$) of the nanocones. The apex angle of the nanofiber approximates to the apex angle of the constituent nanocone, i.e., $\theta = 2\arcsin(\alpha/2\pi)$. In this regard, the effective inner radius and outer radius of the nanofiber (Figure 1b) can be approximated as $r_e = r\sin(\theta/2)$ and $R_e = R\sin(\theta/2)$, respectively. Considering the close packing morphology, i.e., the distance between each layer equals to the graphite distance ($b$ = 3.35 Å), the effective height or length of the nanofiber can be approximated as $h_e = (N-1)b/\sin(\theta/2)$. Thus, the cross-section and volume of the nanofiber can be approximated as $A_e = \pi(R^2 - r^2)\sin(\theta/2)^2$, and $V_e = \pi b(N-1)(R^2 - r^2)\sin(\theta/2)$.

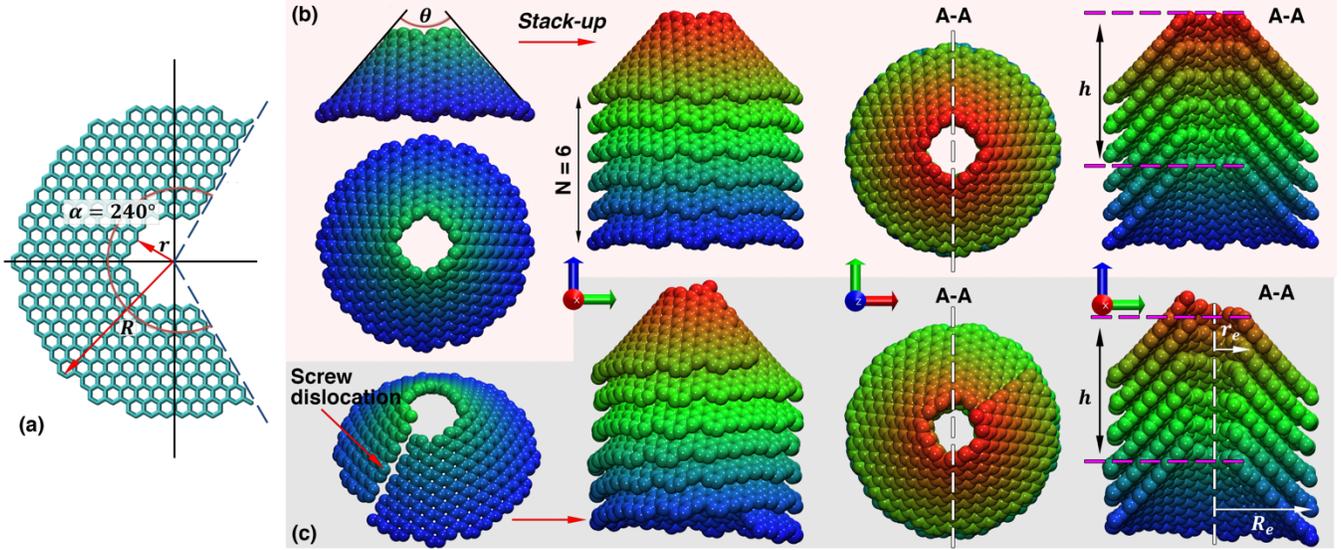

**Figure 1.** Atomic structure of the cup-stacked carbon nanofiber. (a) The fan-shaped graphene nanosheet; (b) The carbon nanofiber constructed from a stacking of nanocones. (b) The carbon nanofiber constructed from a screw dislocation of the nanocones. The atoms are colored according to their z-coordinates along the height direction.

    The atomic interactions between carbon atoms were described by the commonly used adaptive intermolecular reactive empirical bond order (AIREBO) potential [53, 54]. The potential includes short-range interactions, long range vdW interactions and dihedral terms, which has been shown to well represent the binding energy and elastic properties of carbon materials. The vdW interactions were described by the Lennard-Jones term, which has been reported to reasonably capture the vdW in multi-layer graphene [55], multi-wall carbon nanotubes [56], and carbon nanotube bundles [57]. The cut-off distance of the AIREBO potential was chosen as 2.0 Å [58-63]. A small time step of 0.5 fs was used for all calculations with all MD simulations being performed under the software package LAMMPS [64].

    The nanofiber was firstly subjected to energy minimization using conjugate gradient method. According to the atomic configurations, no significant changes were detected for the structure of the nanofiber, and the initial interlayer distance was around 3.30 Å after energy minimization (see Supporting Information S1). Afterwards, the structure was relaxed under isothermal-isobaric (NPT) ensemble using Nosé-Hoover thermostat for 500 ps. Periodic boundary conditions were applied in the axial (or thickness) direction during the energy minimization and relaxation processes. To isolate the thermal influence, a low temperature of 1 K was adopted initially. After relaxation, the structure was switched to non-periodic boundary conditions. Tensile deformation was then performed by applying a constant velocity (i.e., 0.02 Å/ps) to one end of the structure and fixing the other end. Each end contains a whole layer of the nanocone.

During the simulation, the commonly used virial stress was calculated, which is defined as [65]

$$\Pi^{\alpha\beta} = \frac{1}{\Omega}\left\{-\sum_i m_i v_i^\alpha v_i^\beta + \frac{1}{2}\sum_i \sum_{j \neq i} F_{ij}^\alpha r_{ij}^\beta\right\} \quad (1)$$

Here, $\Omega$ is the volume of the system; $m_i$ and $v_i$ are the mass and velocity of atom $i$; $F_{ij}$ and $r_{ij}$ are the force and distance between atoms $i$ and $j$; and the indices $\alpha$ and $\beta$ represent the Cartesian components. Adopting different approaches to calculate the volume of the nanofiber would lead to different absolute stress values but not affecting their relative magnitudes.

## 3. Results and discussion

### 3.1. Super-elastic tensile behavior

Initially, we consider the tensile deformation of NF-S containing six layers of carbon nanocone (including four deformable layers and two boundary layers). The constituent nanocone has an outer and inner radius of about 25.8 Å and 7.4 Å, respectively, and sector angle $\theta = 83.6°$ (or $\alpha = 240°$). For comparison, we also re-investigated the tensile behavior of the counterpart nanofiber with six layers of stacked carbon nanocone without screw discussion (denoted as NF for discussion convenience). **Figure 2**a compares the strain energy curve of the NF-S and NF under tensile deformation. Here, the engineering strain is defined as $\varepsilon = \Delta h/h_0$, where $h_0$ and $\Delta h$ represent the initial length (or height) and the length change of the structure, respectively. Considering the periodic scenario along the axial direction, the initial sample length equals to the equivalent height $h_e$.

Strikingly, we find that the strain energy ($\Delta E$) of the NF-S increases continuously until the tensile strain ($\varepsilon$) approaches an extremely large value of 882%. Afterwards, a zigzag profile of the strain energy is observed, which is similar to the plastic deformation in metallic nanowires [66]. In comparison, its counterpart NF (without screw dislocation) shows a nearly overlapped strain energy profile initially, but experiences full failure at a strain only around 58% (with no further strain energy change, see inset of Figure 2a). Such huge gap indicates the distinct mechanical properties between NF-S and NF as induced by the screw dislocation.

By tracking the virial stress ($\sigma$) in Figure 2b, the whole tensile process can be roughly divided into four stages. In the first stage – the homogeneous deformation stage (I, $0 \leq \varepsilon < 12\%$), the stress increases almost linearly with strain (inset in Figure 2b), and the corresponding strain energy shows a good parabolic relationship with the strain (inset in Figure 2a). The maximum stress is around 1.3 GPa, and the effective tensile modulus approximates 16 GPa (based on the least square fitting with $\varepsilon < 5\%$). Such results are consistent with that reported previously for cup-stacked NF [50]. According to the atomic configurations (Figure 2c), both NF-S and NF deform like a continuum fiber in this stage with no detectable structural changes, which is governed by the interlayer vdW interactions.

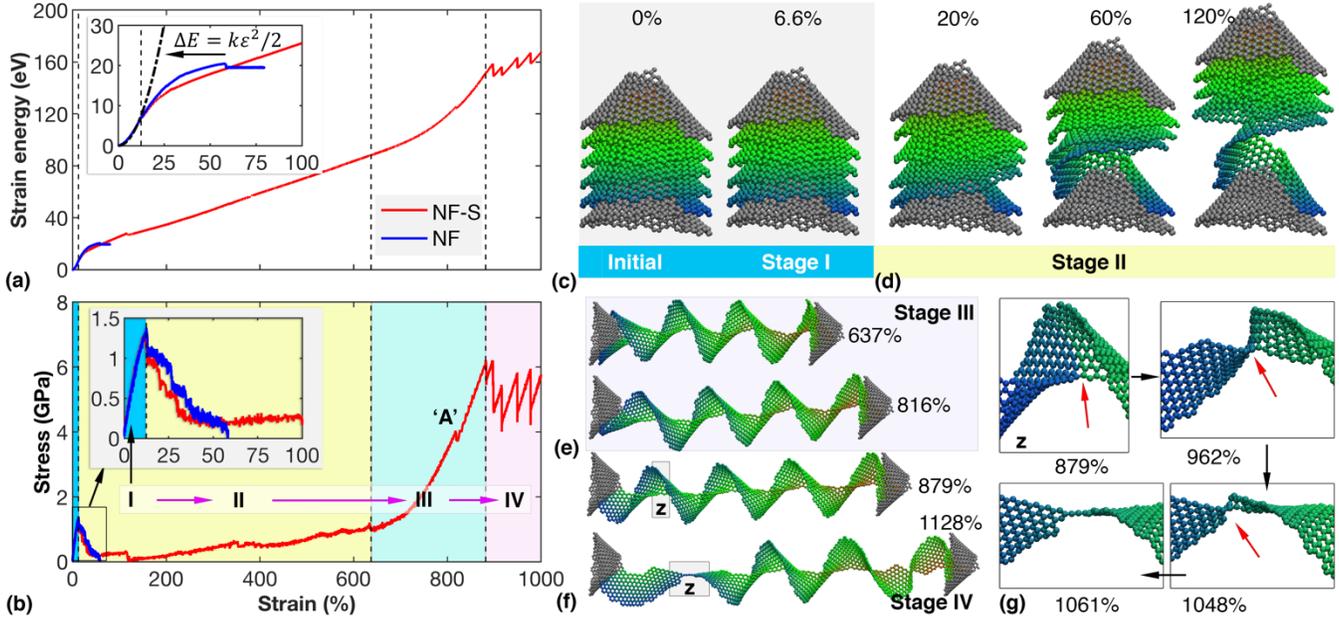

**Figure 2.** Super-elastic tensile behavior of carbon nanofiber. (a) The strain energy as a function of strain. Inset shows the enlarged view of the strain energy profile within the strain of 100%. The dash-dot line is the fitted parabolic profile. (b) The stress-strain curve of the NF-S and NF. Inset shows the enlarged view of the stress-strain curve. I-IV represent the four deformation stages. The atomic configurations at the deformation stage of: (c) I; (d) II; (e) III; and (f) IV. (g) The local atomic configurations showing the bond breaking and formation process of the monoatomic chain at the deformation region z in (f). The red arrows highlight the fracture region in the nanocone. For figures e to g, atoms are colored according to their z-coordinates along the height direction. The grey colored layers in the two ends are the rigid boundary layers.

After stage I, the tensile stress decreases continuously to zero for the NF (due to the full separation of nanocones). For the NF-S, although the stress exhibits an initially similar decreasing pattern, it increases gradually until a large tensile strain of ~637%. According to Figure 2d, the nanofibers undergo a delamination process (stage II), which occurs in multiple places initially (in both NF and NF-S), and then concentrate on one site with further tensile deformation (see Supporting Information S2 for more details). Such fact leads to local fluctuations to the strain energy and stress profile as shown in Figure 2a and 2b, respectively. From Figure 2a, the strain energy of the NF-S increases almost linearly with the tensile strain, which indicates that the tensile force is a constant during the deformation in stage II. Such observation can be explained from the perspective of the vdW interlayer interactions. It is expected that although the nanocone of the NF-S experiences bending deformation, the overall deformation of the NF-S is dominated by interlayer delamination. In this regard, considering the free surface energy density as $\gamma_s$, the stable delamination process will linearly create new free surface, and thus leads to a dominant linear increase of the strain energy ($\Delta E \approx A_{free}\gamma_s$).

After the delamination stage (II), the tensile stress of the NF-S starts to increase significantly until the strain reaches around 882%. According to the atomic configurations in Figure 2e, stage III is a combination of further delamination, and strong distortion and stretching of covalent carbon bonds. At the strain of 637% (after stage II), the delamination of the layer adjacent to the bottom end does not occur. With further increase of the tensile stress, further delamination is observed, which is completed at the strain of 816%. Such facts result in a nonlinear increasing stress (before the inflection point "A" in Figure 2b). After the full delamination, the deformation of NF-S is dominated by the distortion and stretching of covalent bonds, which thus leads to a linear stress increasing profile. The maximum stress is about 6.2 GPa. In stage IV, the bond failure is initiated at the armchair corner of the inner edge (Figure 2g), and the increasing stretch further tears apart adjacent nanosheet at the fracture region, which eventually leads to the formation of monoatomic chain (Figure 2f). Specifically, the zigzag profile is resulted from the occurrence of bond breaking event.

To exploit the tensile behavior of the NF-S, we examine the recovery capability of the deformed structure through energy minimization simulation. For such purpose, the external constant velocity is removed and the axial degree of freedom of the loaded end is released. The other end of the structure remains fixed during the energy minimization, and the energy minimization is performed for the NF-S with different tensile strains separately. **Figure 3** compares the potential energy change (including the vdW interlayer interactions and covalent bond energy – the REBO term in the potential) during tensile and energy minimization process. As illustrated in Figure 3a, the vdW interactions increase significantly during the tensile deformation stages I and II, which is in line with the delamination process. In comparison, the covalent bond energy change is much smaller and experiences a gradual increase during this period of deformation, which increases significantly in the deformation stage III (Figure 3b). During the energy minimization, both vdW energy and covalent bond energy differences are well overlapped with that under tensile deformation, which restores to zero at the end of the minimization, suggesting an elastic recovery process.

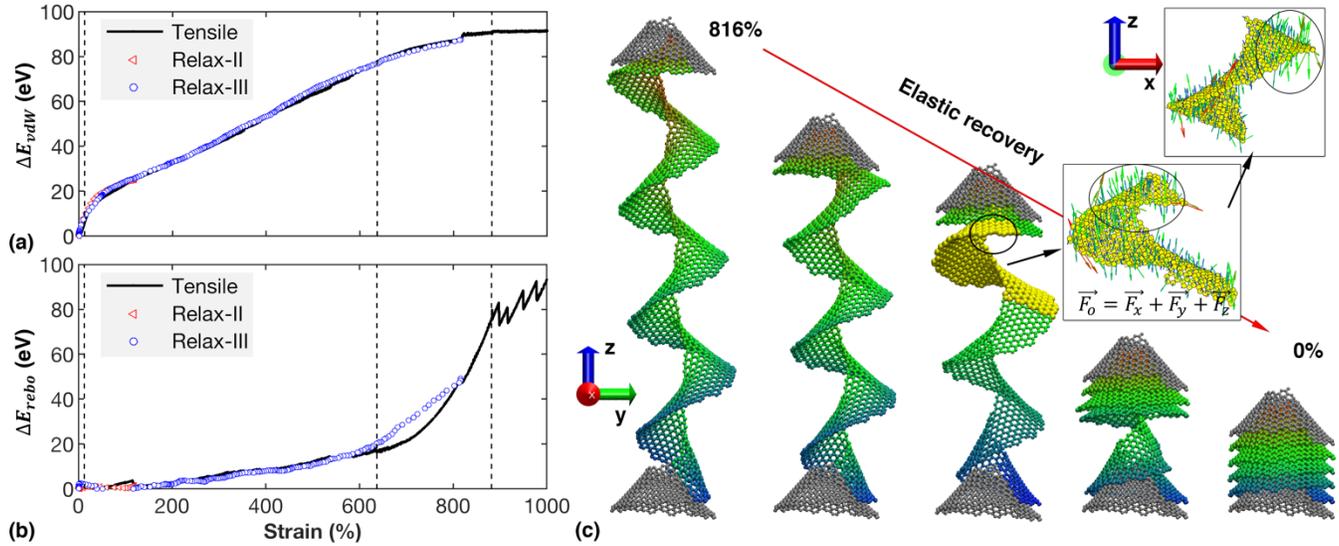

**Figure 3.** Elastic recovery of the stretched carbon nanofiber. (a) The change of vdW interactions during tensile and energy minimization process. Relax-II and Relax-III represent the relaxation from elastic deformation state II and III, respectively. (b) The change of covalent bond energy (REBO term in the potential) during tensile and energy minimization process. (c) Atomic configurations showing the full elastic recovery of the structure from stage III. Inset shows the force vectors for the atoms in yellow color. The other atomic structures are colored according to the atomic coordinates in the height direction. The layers in the two ends as colored in grey are the boundary layers.

Tracking the atomic configurations, the stretched NF-S resumes to its initial configuration with no observable changes (Figure 3c). Such results signify that the deformation stages of I, II and III are all elastic deformations. The elastic recovery in stage I and II is triggered dominatingly by the interlayer vdW interactions. While, the elastic recovery in stage III is accomplished by the combination of vdW interactions and the covalent bond recovery. As shown in the insets of Figure 3c, larger atomic forces are observed in the adjacent overlapping regions, which drives the elastic recovery. It is noted from Figure 3c that the recovery is dominated by the top end of the structure.

To affirm the intriguing tensile characteristic of the NF-S, we also investigate the strain rate influence. Firstly, we perform the relaxation simulation by holding the loaded end of the structure at different strains (selected randomly from the three elastic deformation stages), mimicking a quasi-static tensile deformation. As compared in **Figure 4**a, the potential energy of the stretched NF-S remains almost unchanged during the relaxation. According to the atomic configurations, the stretched NF-S undergoes free vibration with a potential energy change less than 0.01% (see insets in Figure 4a). Secondly, we apply different velocities to stretch the NF-S, including 0.02 Å/ps, 0.10 Å/ps, 0.20 Å/ps, 0.30 Å/ps, and 0.40 Å/ps. As compared in Figure 4b, the profiles of strain energy obtained from different strain rates are nearly overlapped with each other. No observable differences are found in the atomic configurations of the NF-S under different strain rates. These results signify that the super-elastic tensile properties of the NF-S are not affected by the strain rate.

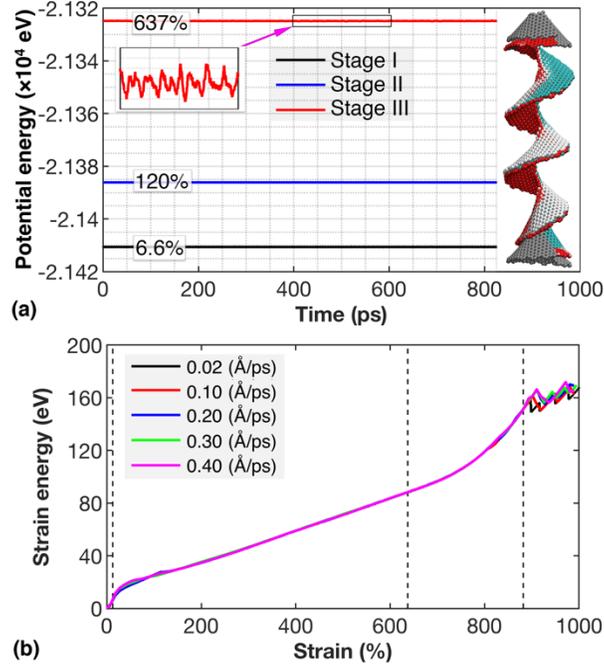

**Figure 4.** Strain rate influence. (a) The potential energy of the stretched NF-S during relaxation. The three strains are picked from the three elastic deformation stages. Inset A shows the potential energy fluctuations due to the free vibration of the structure. Inset B illustrates the atomic configurations at the simulation time of 0 ps (red), 250 ps (white), and 625 ps (cyan), showing the free vibration. The layers colored by gray at the two ends are the boundary layers. (b) The strain energy of the NF-S under different strain rates.

### *3.2. Geometrical influence*

It is of great interest to know how the geometrical parameters will impact the super-elastic behavior of the NF-S, including the layer number ($N$), inner ($r$) and outer ($R$) radius, and the apex angle ($\theta$). For such purpose, we first analysis the tensile force during the deformation as shown in **Figure 5**. Basically, the force shares the similar changing pattern as that of the tensile stress in Figure 2b. Here, we focus on the critical force $F_i$ and strain $\varepsilon_i$ that are associated with each deformation stage ($i$ = I, II, and III, respectively). In stage I, the force increases significantly until reaching a critical value ($F_I$, about 9.4 nN), which is the largest magnitude during the whole tensile process. After passing this threshold value, the NF-S undergoes a delamination-dominated deformation process (in stage II), which is analogous to a peel process. Comparing with stage I, the delamination occurs only between a small area of the adjacent layers, and thus only a small amount of force is desired to continue the delamination. As a consequence, the force decreases firstly and then fluctuates around a small value ($F_{II}$, about 1.3 ± 0.3 nN) during the whole deformation stage II. After stage II, the force starts to increase again due to the distortion and stretching of covalent carbon bonds. The maximum force in stage III ($F_{III}$) is dominated by the strength of the carbon bonds. According to the deformation process in Figure 2c to 2f, $\varepsilon_I$ is determined by the vdW interactions, $\varepsilon_{II}$ is related with the delamination process, and $\varepsilon_{III}$ is determined by the carbon bonds (which is also the yielding strain of the NF-S).

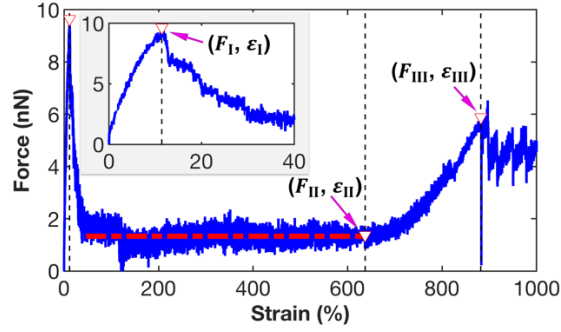

**Figure 5.** Tensile force of the NF-S during the deformation. The dash-dot line represents the averaged force in stage II. Insets show the force profile at the strain range between 0 and 40%.

With above understanding, we then assess the influences from the four geometrical parameters separately. Four groups of tensile simulations are carried out for the NF-S with varying geometrical parameters (following the same simulation settings) as summarized in Table 1. The first group shows that the apex angle remarkably affects the tensile behavior of the NF-S. As illustrated in **Figure 6**a, the NF-S with the largest apex angle (112.9°) exhibits an extraordinarily large elastic limit ($\varepsilon_{III}$, up to 1289%). However, there is almost no delamination occurring for the structure with a small apex angle of 38.9°, which results in a much smaller elastic limit of about 99%. Note that the smallest apex angle of 19.2° is not considered as the selected inner radius is too small to form a screw dislocation. Recall the atomic configurations in Figure 2e, the delamination stage is essentially regulated by the inner edge of the NF-S. For the structure with same inner radius but smaller apex angle, the initial unit height (i.e., $h_e^u = b/\sin(\theta/2)$) will be much larger. In other words, the available stretching space has been significantly suppressed, and thus leads to the diminishing of the delamination process. For illustration, insets of Figure 6a show the atomic strain within the NF-S with $\theta = 38.9°$, from which the deformation is majorly occurring around the inner edge. Here, the atomic strain is just a placeholder for the continuum strain, which is defined as $\varepsilon_a = \sum_M \varepsilon_i^b /M$, where, $\varepsilon^b = (b_d - b_0)/b_0$ is the bond length changing ratio, and $b_0$ and $b_d$ represent the initial and deformed bond length, respectively. $M$ is the number of bonds for the corresponding atom.

**Table 1.** Geometrical parameters of the NF-S. V representing the varying parameter. *N* is the layer number; $R_e$ and $r_e$ are the outer and inner radius, respectively; $\theta$ is the apex angle of the corresponding nanocone.

| Group | N | $R_e$ (Å) | $r_e$ (Å) | $\theta$ (°) |
|---|---|---|---|---|
| 1 | 6 | 25.8 | 7.4 | V |
| 2 | 6 | 25.8 | V | 83.6 |
| 3 | 6 | V | 7.4 | 83.6 |
| 4 | V | 25.8 | 7.4 | 83.6 |

Figure 6b compares the overall influence on the critical forces of the NF-S from the apex angle. Basically, $F_I$ increases significantly with the apex angle, which is reasonable due to several factors. For instance, smaller apex angle will lead to less overlapping areas (less vdW interactions) between adjacent layers and their relative movement becomes more like sliding during tensile deformation, and thus it will become much easier to trigger the delamination. In comparison, the apex angle exerts minor influence on $F_{II}$ and $F_{III}$. Considering that $F_{II}$ is determined by the delamination front area and the width ($w = R - r$) of the nanocone is the same, it is thus reasonable to find a minor influence from the apex angle. As aforementioned, $F_{III}$ is determined by the covalent bond strength and the deformation is mainly concentrated at the inner edge of the structure, therefore, it is understandable that the apex angle exerts ignorable influence on $F_{III}$. In terms of the critical strain, totally different influential pattern is observed (Figure 6c). In detail, $\varepsilon_I$ shows minor variation due to the change of the apex angle. However, due to the suppressed delamination process at lower apex angle, both $\varepsilon_{II}$ and $\varepsilon_{III}$ show a significant reduction when the apex angle decreases. It is interesting to note that the difference between $\varepsilon_{II}$ and $\varepsilon_{III}$ ($\Delta\varepsilon = \varepsilon_{III} - \varepsilon_{II}$) is nearly a constant at different apex angles, implying that the apex angle exerts ignorable influence on the elastic deformation stage III.

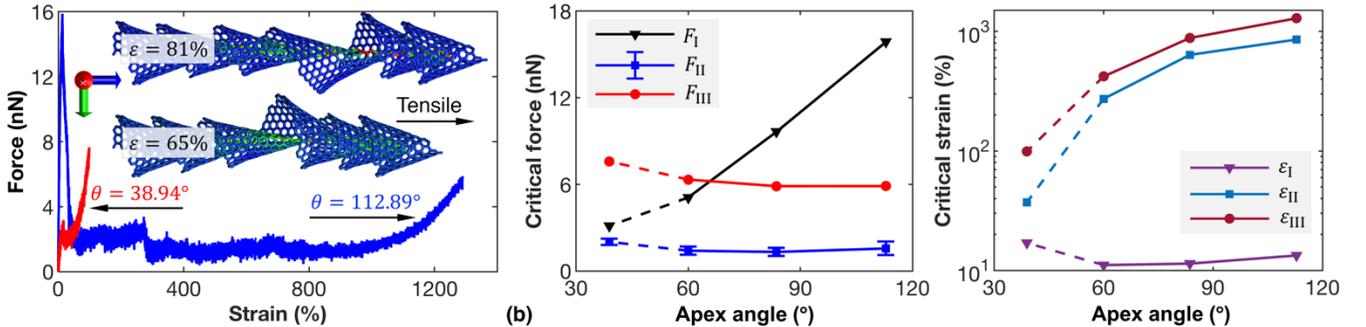

**Figure 6.** Tensile deformation of NF-S with different apex angles. (a) The tensile force as a function of strain (only the elastic deformation results are plotted). Insets show the atomic configurations of the NF-S with an apex angle of 38.9°. Atoms are colored according to their strain between -1% (blue) and 8% (red).

(b) The critical forces as a function of apex angle. (c) The critical strains as a function of apex angle.

**Figure 7**a shows the tensile force as a function of strain for the NF-S with different inner radius (from 7.4 Å to 14.8 Å) but the same outer radius of 25.8 Å (Group 2). It is found that, the NF-S with larger inner radius has a smaller $F_I$ and $F_{II}$, but $F_{III}$ is almost the same around 6 nN. In terms of the critical strain, $\varepsilon_I$ is nearly the same for the NF-S with different inner radiuses. However, $\varepsilon_{II}$ increases remarkably when the inner radius increases, which consequently leads to an increasing $\varepsilon_{III}$. For instance, the NF-S with an inner radius of 7.4 Å has a $\varepsilon_{III}$ of about 882.0%, which increases to 1727.8% when the inner radius increases to 14.8 Å. Comparing with $\varepsilon_I$ (~ 11.6%), $\varepsilon_{III}$ is around 150 times higher. These results suggest that the inner radius can be used to tailor the elastic limits of the NF-S.

According to Figure 7b, the outer radius induces different influences on the tensile behavior of the NF-S (Group 3). Specifically, the critical forces in all three elastic deformation regimes increase when the outer radius increases, however, the critical strains are almost unchanged. Refer to the atomic structure (Figure 1a), the influences from the inner or outer radius can be explained from the perspective of the width of the nanocone. As summarized in Figure 7c, the width exerts a uniform impact on the critical forces. In particular, $F_I$ and $F_{II}$ increase almost linearly with the increase of the width, which is reasonable as larger width leads to more contact area between adjacent layers and thus stronger interlayer vdW interactions. On the other hand, $F_{III}$ fluctuates around 6.2 nN for the NF-S with varying inner radius but exhibits a gradual increasing tendency initially and saturates around 6.2 nN for the structure with varying outer radius. Such gradual increasing tendency is related with the stress or strain distribution along the inner edge of the NF-S in stage III. For the structure with the same inner radius, but smaller outer radius, the narrower ribbon structure has been severely bent during the tensile deformation, which leads to a much larger strain at the end (panel A in Figure 7e) compared with other sections (panel B in Figure 7e). Such fact results in easier failure. According to the atomic configurations in Figure 7f, the wider ribbon can maintain the vdW interactions at the edge during the tensile deformation, which helps to release the local strain at the edge. As a result, the tensile strain of the inner edge is better distributed (panel A in Figure 7f), and thus the structure can bear larger tensile force before fracture. It is worth noting that for all studied NF-S, the strain at the armchair edge is higher than that at the zigzag edge, which is in line with that observed during the tensile deformation of graphene [67].

Comparing with the critical force, the critical strain changes differently with the width when the inner or outer radius changes. From Figure 7d, the critical strain in stage I ($\varepsilon_I$) fluctuates around 12% for the NF-S with varying inner radius but increases from 8.9% and saturates around 15.2% for the counterpart with varying outer radius. In combination with the higher critical force $F_I$, these results suggest that larger outer radius can enhance the resistance of the structure to delamination under stretch. The most significant impacts from the inner edge is found on the elastic regime II as illustrated in Figure 7a. Since some NF-S exhibits a mixed deformation mode in stage III, the determination of $\varepsilon_{II}$ suffers from some uncertainty, we thus focus on the changing tendency of $\varepsilon_{III}$. According to Figure 7d, $\varepsilon_{III}$ reduces linearly and remarkably with the inner radius. As the deformation is concentrated at the inner edge of the structure in stage III, it is thus expected that the difference between $\varepsilon_{II}$ and $\varepsilon_{III}$ is a constant (i.e., $\Delta\varepsilon_{II-III} = C$) similar as observed from the structure with different apex angles. Recall the atomic configurations (Figure 7e and 7f), the structure starts to enter elastic deformation stage III when the inner edge is approaching the fully stretched status from the helical morphology. In other words, considering the equivalent inner edge length of a unit cell as $L_e^u$ and the initial height of $h_e^u$, the critical strain $\varepsilon_{II}$ can be estimated from $\varepsilon_{II} = (N_{eff}L_e^u - N_{eff}h_e^u)/N_{eff}h_e^u$, here $N_{eff}$ represents the effective layer number (excluding the two boundaries). Thus, the critical strain can be estimated from $\varepsilon_{III} = (L_e^u - h_e^u)/h_e^u + C$, and the effective inner edge length can be estimated from $L_e^u \approx 2\pi r$. To note that, due to the armchair edge at each corner of the hexagonal morphology, the actual $L_e^u$ should be slightly larger. Therefore, $\varepsilon_{III} \approx \frac{2\pi \sin(\theta/2)}{b}r + C - 1$. It is evident that $\varepsilon_{III}$ depends linearly on the inner edge radius $r$, which fits well with the MD results in Figure 7d.

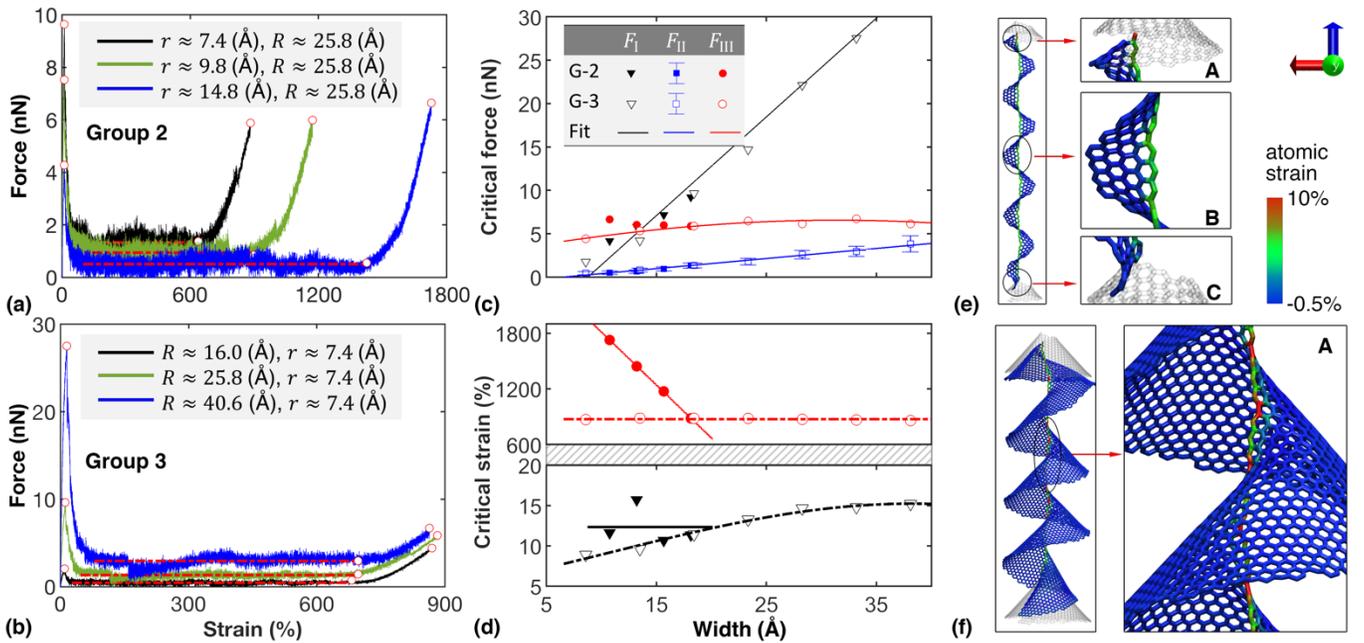

**Figure 7.** Tensile deformation of NF-S with different width. (a) The tensile force as a function of strain (only the elastic deformation results are plotted) for the NF-S with varying inner radius. (b) The tensile force as a function of strain (only the elastic deformation results are plotted) for the NF-S with varying outer radius. (c) The critical forces as a function of width. (d) The critical strains as a function of width. The atomic configurations showing the atomic strain distribution at the tensile strain of 515% for the: (e) NF-S with an inner radius of 7.4 Å and outer radius of 16 Å; (f) NF-S with an inner radius of 7.4 Å and outer radius of 35.7 Å. The layers colored by gray at the two ends are the boundary layers.

**Figure 8** shows the influence on the tensile properties of the NF-S from the layer number (Group 4). As is seen, the layer number exerts insignificant influence on the first and second elastic deformation stage, but affects the elastic deformation stage III. Specifically, $F_I$ is found to decreases initially and then saturates around 7.5 nN when the layer number increases, indicating the NF-S with larger layer number is easier to enter the delamination deformations stage II. On the other hand, $F_{II}$ and $F_{III}$ fluctuate around 1.2 nN and 6 nN, respectively (Figure 8b). Interestingly, the critical strain exhibits a length/height dependency. All critical strains show a gradual decreasing tendency with the increasing layer number, which saturates around a certain value when the layer number is over 10. The smaller critical strain $\varepsilon_{III}$ signifies that the NF-S with higher layer number is easier to break.

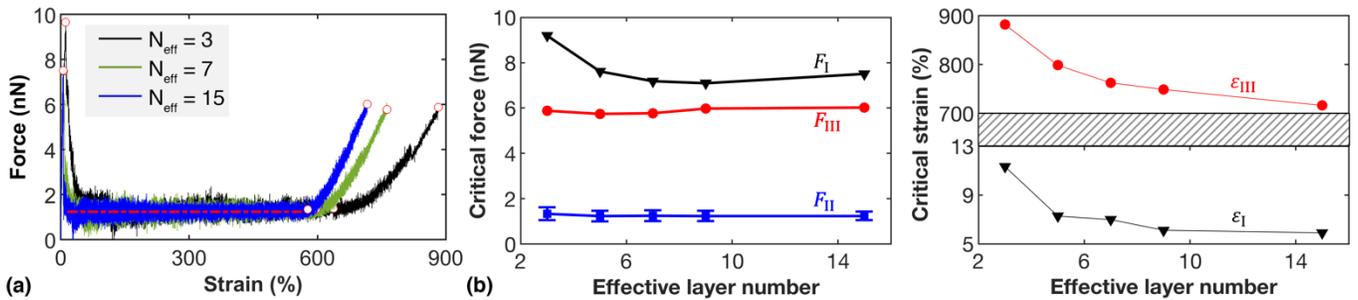

**Figure 8.** Tensile deformation of NF-S with different layer numbers. (a) The tensile force as a function of strain (only the elastic deformation results are plotted). (b) The critical forces as a function of the effective layer number. (c) The critical strains as a function of the effective layer number.

### *3.3. Compressive behaviours*

Besides the tensile deformation, it is also of great interest to explore the compressive behaviours of the NF-S. Following the tensile simulation, we first consider the compressive behaviour of the NF-S with one end fixed and the other end being compressed by a constant velocity of 0.02 Å/ps. As shown in **Figure 9**a, the structure deforms elastically until the strain reaches 32.7%. According to the atomic configurations, the inner edges experiences the most severe compressive deformation, and interlayer bonds are formed when the strain is larger than 32.7% (Figure 9b). We also examine the NF without the screw dislocation, from which a similar deformation process is observed and the strain energy profile is almost overlapped with that from the NF-S (Figure 9a).

Above results suggest that NF-S and NF share similar compressive deformation behavior when both ends are fixed. Considering the cone shape, we further test the deformation scenario by only fixing one end of the sturcture. That is a virtual wall is introduced at the bottom of the sample, and the NF-S is compressed by a constant velocity applied on the top of the structure (left of Figure 9c). The virtual wall will exert a repulsive force to the atoms approaching the wall, which is expressed by $F = -kd^2$. Here, $k$ is the force constant and $d$ is the distance from the atom to the wall. As plotted in Figure 9a, the NF-S shows a totally different strain energy profile. Comparing with the above scenario with both-ends fixed, the NF-S shows a much larger elastic regime but a much smaller strain energy magnitude. According to the atomic configurations in Figure 9c, the NF-S deforms through a flatten and sliding process, i.e., the cone shape is firstly flattened during the compression and then slides away from the compression zone. Such deformation process leads to a dominant tensile strain within the nanosheet rather than compressive strain as seen in the right panel of Figure 9b. Similarly, the NF shows a flatten deformation process. However, due to the lacking of screw dislocation, the flatten deformation is only limited at the outer edges of the cone shape and no sliding process can be triggered. As shown in Figure 9d, the C-C bonds at the outer edges have been severely stretched. Overall, these results suggest that the presence of screw dislocation enables the NF-S with intriguing deformation features.

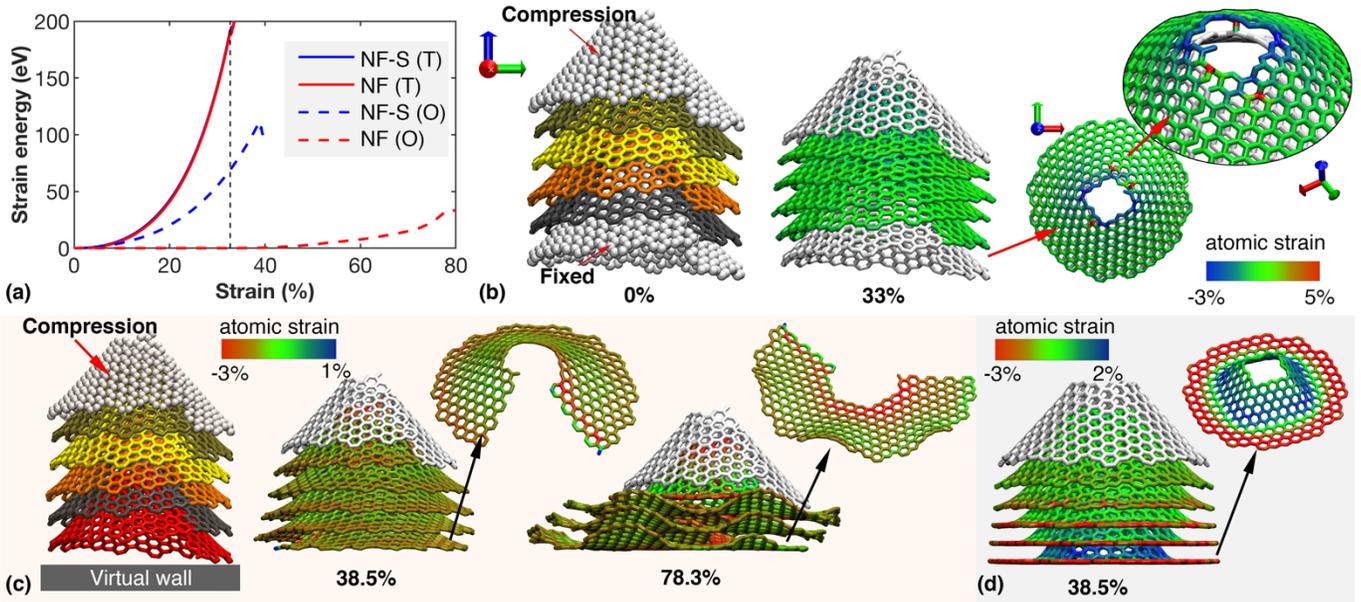

**Figure 9.** Compressive deformation of NF-S under different simulation settings. (a) The strain energy as a function of strain. T and O represent the two-end fixed and one-end fixed simulation scenario, respectively. (b) The atomic configurations showing the deformation process of the NF-S when both ends are fixed. (c) The atomic configurations showing the deformation process of the NF-S when only one end is fixed. The virtual wall imposes a repulsive force on the approaching atoms. (d) The atomic configurations showing the deformation process of the NF when only one end is fixed. The atomic configurations in the left panel of b and c are colored according to the layer number, other atomic configurations are colored according to the atomic strain. The white layers at the end of the structure are the boundary layers.

It is worth noting that our simulations are performed for a low temperature of 1 K. This is also commonly applied in literature when investigating the mechanical properties of nanomaterials in order to remove the thermal influence. To further affirm the super-elastic properties of the NF-S, we extend the tensile simulations under a temperature of 100 K, 200 K, and 300 K. As compared in **Figure 10**, the strain energy profiles are almost overlapped with each other, indicating a same super-elastic tensile deformation characteristic. In detail, the temperature shows insignificant influence on the elastic deformation regime I and II. In stage III, the yielding strain $\varepsilon_{III}$ (as well as the critical force $F_{III}$) decreases slightly when the temperature increases. Such observation agrees with other nanostructures being reported, such as the carbon nanothreads [68], graphene and graphyne [69]. We should also note that current work has focused on a perfect structure with unsaturated inner and outer edges. It is expected that different edge conditions (like the full H-termination or partial H-termination), the presences of different defects (such as vacancies or substitutional atoms), and interlayer covalent bonds would affect the performance of the nanofiber, which will be explored in our coming work.

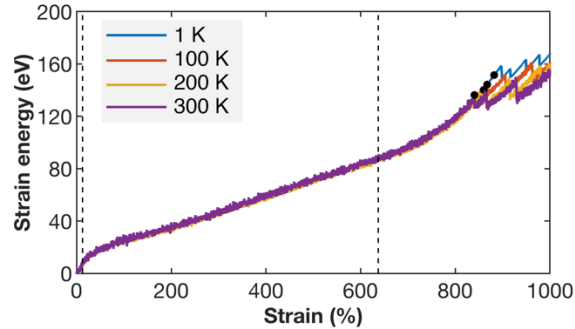

**Figure 10.** Temperature influence on the tensile properties of the NF-S. The circle markers highlight the critical strain in elastic deformation stage III.

## 4. Conclusions

Based on the large-scale molecular dynamics simulations, this work explores the tensile characteristics of carbon nanofibers (NF-S) constructed from a screw dislocation of carbon nanocones. It is found that NF-S possesses a super-elastic tensile property with the yielding strain exceeding 1000%, which is realized by the delamination process of the nanocones. Specifically, the elastic deformation of NF-S can be divided into three stages, including the initial homogeneous deformation, the delamination of the nanocones, and the further tensile deformation due to the stretching and distortion of covalent C-C bonds. The initial homogeneous tensile deformation is determined by the interlayer vdW interaction strength, and therefore NF-S and NF share the same deformation in this stage, i.e., an overlapping strain energy profile. Atomic configurations show that the third elastic deformation stage is dominated by the inner edges of the nanocones, where a strong strain concentration is observed during tensile deformation. Different strain rates and temperatures have been tested, from which a uniform super-elastic tensile deformation is observed.

The geometrical parameters including the apex angle, inner and outer radius, and the layer number are found to exert different influences on the tensile properties of the NF-S. In detail, the apex angle exerts a remarkable impact on the delamination stage, and a larger apex angle induces a larger delamination regime and thus a higher elastic limit. Meanwhile, the force required to trigger the delamination is larger for the NF-S with higher apex angle. The inner and outer radius exert different impacts on the critical strains and critical forces. Specifically, the force to trigger the delamination increases when the inner radius decreases or outer radius increases, which is originated from the width changes of the nanocone. A wider width means stronger interlayer interactions and thus requires higher force to delaminate. On the other hand, the delamination regime is dominated by the inner radius of the nanocone, i.e., the allowable delamination increases almost linearly with the increase of inner radius. It is further found that the yielding strain decreases initially with the increase of layer number, and then saturates around a contain value. Meanwhile, the critical force to trigger the delamination shares the same changing tendency. For all the examined structures, the tensile force at the yielding point where bond failure is observed, is nearly a constant (~ 6 nN). Such observation is consistent with the fact that the deformation is concentrated at the inner edges of the nanocone.

Additionally, it is found that the compressive behaviour of the NF-S is highly dependent on the loading scenarios, and the screw dislocation allows the nanocones to be flattened and slide away from the compression regime. In summary, this study provides a first-time understanding of the mechanical properties of NFs with a screw dislocations of carbon nanocones, which should shed lights on their novel applications, such as nanoscale springs or nanoscale actuators.

**Supporting Information**

The Supporting Information is available free of charge, including: the relaxed structure of the carbon nanofiber; and the atomic configurations during tensile deformation.

**AUTHOR INFORMATION**

**Corresponding Author**

*E-mail: duanshuyong@hebut.edu.cn; zhan.haifei@qut.edu.au*E-mail: duanshuyong@hebut.edu.cn; zhan.haifei@qut.edu.au

**Author Contributions**

XH and FX contribute equally to the work. FX and HZ carried out the simulations. XH, FX, HZ, YG, SD, and GL conducted the analysis and discussion.

**Notes**

The authors declare no competing financial interests.

**ACKNOWLEDGEMENT**

Support from the Youth Program of National Science of China (Grant No. 51805141) and Hebei Natural Science Foundation of Youth Science Foundation (Grant No. E2018202243) are gratefully acknowledged (SD). Support from the ARC Discovery Project (DP170102861) and the High-Performance Computing (HPC) resources provided by the Queensland University of Technology are gratefully acknowledged (HZ, YG). HZ would also like to acknowledge the support from the Start-up Fund from Queensland University of Technology.